\documentclass[%
 reprint,
 amsmath,amssymb,
 aps,
prb
]{revtex4-2}

\usepackage{graphicx}
\usepackage{dcolumn}
\usepackage{bm}

\begin{document}

\preprint{}

\title{Understanding the lifetime of water with dynamic network analysis: the case of CsOH$\cdot$H$_2$O}
\author{Graeme J. Ackland}
\author{Ciprian G. Pruteanu}
\author{John S. Loveday}
\author{Keishiro Yamashita}
 \affiliation{School of Physics and Astronomy, University of Edinburgh.}

\date{\today}

\begin{abstract}

We describe the atomic-level motions in caesium hydroxide monohydrate (CsOH$\cdot$H$_2$O), which is a chemical compound containing layers of water and hydroxide ions.  At this composition, each oxygen is involved in three hydrogen bonds which, in the hexagonal structure, form a quasi-2D honeycomb lattice.   While oxygen and caesium atoms form a typical crystal lattice, the dynamics of the hydrogen atoms are more complex.  Here we show that the covalent and hydrogen bonds are continually interconverting, meaning that the water and hydroxyl are interconverting by proton exchange.  The order-disorder transition of the water and hydroxyl proceeds by chemical reaction rather than rotation or diffusion of the molecules.  A hydrogen can rotate out of the layer, leaving a vacant site in the 2D layer. Such a hydrogen vacancy can diffuse rapidly by single molecule rotation, leading to fast-ionic conduction.
The proton exchange leads to a novel type of Raman activity combining stretch and exchange processes, for which we develop a theoretical model. This would manifest in a broad single peak associated with both H$_2$O and OH stretches and a low frequency peak appearing at elevated temperature. 
\end{abstract}

\maketitle


\section{\label{sec:level1}Introduction:}

Consider the classic undergraduate physics question  ``How many molecules of water from  Socrates' cup of hemlock are in my tea?"\cite{shapley1964stars}.  The answer, several thousand,  illustrates the vastness of Avogadro's number compared to the oceans measured in cups.  Among the approximations required is that
the water molecules are indestructible.  If this is not true, there are N$_A^2$ permutations of hydrogens among the N$_A$ molecules - once the hydrogens exchange, there are none of the same molecules from your own cup of tea present\footnote{Typically about 10$^{25}$ molecules in the cup, 10$^{22}$ cups in the oceans, so the standard answer is about 10$^3$.  Given those  10$^{47}$ molecules on earth, if the hydrogen mixes the chance of a given oxygen having the same two hydrogens is 10$^{-94}$.  So that means of order 10$^{-69}$  molecules in both cups.    Which, OK, isn't  exactly zero}.
Beyond this whimsy, there are many practical reasons to understand how easily water molecules break and reform.

CsOH$\cdot$H$_2$O is a hydrated alkali hydroxide: one caesium hydroxide unit plus one water molecule.
It is the simplest hydrate of CsOH, with both ionic and hydrogen bonding\cite{Hermann2016}.   Caesium Cs$^+$ ions are arranged in a triangular layer, and the OH$^-$ and H$_2$O form a nearly planar honeycomb hydrogen-bonded layer 
in the trigonal $P3m1$ \cite{jacobs1982struktur,marx1990time} structure, or a 3D hydrogen-bonded network in a tetragonal $I4_1/amd$ \cite{vcerny2002tetragonal} structures.
It was recently reported to be a  ``superprotonic conductor"\cite{rodenburg2025superprotonic} which would imply a $P6/mmm$ symmetry for the 2D layered phase. 

The structures have Cs–O distances around 3.2–3.4\AA\, (due to large Cs$^+$ radius).  O–H distances are around 1 \AA{} in both OH and water molecules.  The hydrogen bonds: O–H···O distances are 1.8–2.0 \AA.  The tetragonal structure forms a 3D hydrogen bonded network, while the hexagonal one is 2D.

The honeycomb layers can be regarded as a 2D analogue of water, and caesium hydroxide hydrate is a model system for studying hydrogen bonded networks.  Specifically, it is assumed that the hydrogens are disordered but obey the ice rules with precisely one hydrogen between each pair of oxygen atoms, bonded to one or other subject to those oxygens have one or two bonds. This requires three oxygen neighbours per oxygen, which leads naturally to the honeycomb structure.    

The ice rules place only weak, long ranged constraints on the arrangement of water molecules and hydroxide molecules e.g. it is allowed to have a six-membered ring comprised entirely of OH$^-$ or H$_2$O, but the neighbouring ring must contain at least one of the other species (Fig.\ref{fig:layer}).   

At room temperature it is observed that the water and hydroxide molecules are positionally ordered, i.e. alternately on the lattice, while the hydrogens are disordered i.e. bonded to one or other of their neighbouring oxygens, constrained by the ice rules.  At low temperatures, there should be an ordered phase, suggested by the emergence of optical modes in Raman and IR spectroscopy \cite{Lutz1988} as well as the monoclinic distortion \cite{jacobs1982struktur} while at high temperatures there should be an OH-H$_2$O disordered phase prior to melting \cite{marx1990time}.  The kinetics of such a transition are unknown - plausible candidates are the diffusion of OH and H$_2$O units or proton exchange\cite{stahn1983}.  The latter mechanism involves a so-called identity reaction in which the reagents and products are chemically identical.  This provides a contrast with ice. Combined with the 2D nature of the network this makes CsOH$\cdot$H$_2$O a simpler system to study proton dynamics in a hydrogen-bonded network .

Equating experimental and calculation symmetry in this system requires care, because the structures reported from diffraction are ``disordered".  This means that, in a calculation where the hydrogen positions must be specified, enforcing the reported symmetry leads to high-energy structures with imaginary phonons.  In almost all cases, these  structures contain hydrogens centred between two oxygens, a chemically nonsensical arrangement.  When the symmetry is broken, OH and H$_2$O molecules form and the calculated energy drops considerably.   These structures typically have lower symmetry (often monoclinic Cm, Pm Pc, or even $P1$) than the trigonal symmetry ($P3m1$) from the experiment \cite{marx1990time}.

\section{Methods}

\subsection{DFT Calculations}

We carried of ab initio molecular dynamics using CASTEP 25.11\cite{clark2005first} at a range of temperatures.  The calculations used a 180 atom unit cell with a vicinal orientation, creating a single 5x6 honeycomb layer with 30 oxygens.  This geometry allows us to avoid short repeat distances through the periodic boundaries.  We used the more accurate rSCAN functionals\cite{bartok2019regularized} with on-the-fly ultrasoft pseudopotentials for structural relaxations, where we are interested in small enthalpy differences, and the simpler PBE\cite{perdew1996generalized} for the DFPT and molecular dynamics where we are interested in dynamics.

We also used CASTEP to examine static structures and calculate Raman spectra with the default DFPT method.  As we shall see, perturbation theory is inappropriate for describing the highly anharmonic motion of the hydrogen atoms, which in many cases occupy a double-well potential.

\subsection{Elementary crystal structures}

Various structures have been found and proposed for CsOH$\cdot$H$_2$O (Monohydrate).  Notably the tetragonal $I4_1/amd$ reported by $\mathrm{\check{C}ern\acute{y}}$\cite{vcerny2002tetragonal} and the nearly-hexagonal  $P3m1$ and  $P6/mmm$  structures  examined here.

\subsubsection{Tetragonal structure - 3D network}

\begin{figure}
\includegraphics[width=\linewidth]{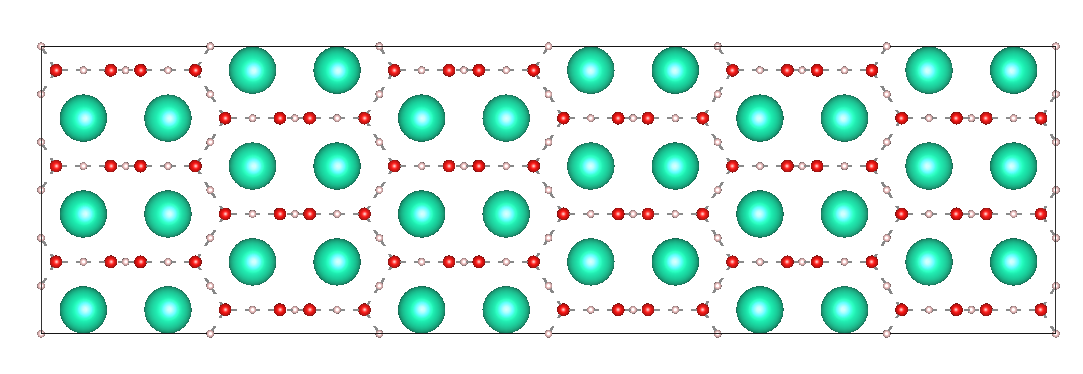} \\
\includegraphics[width=\linewidth]{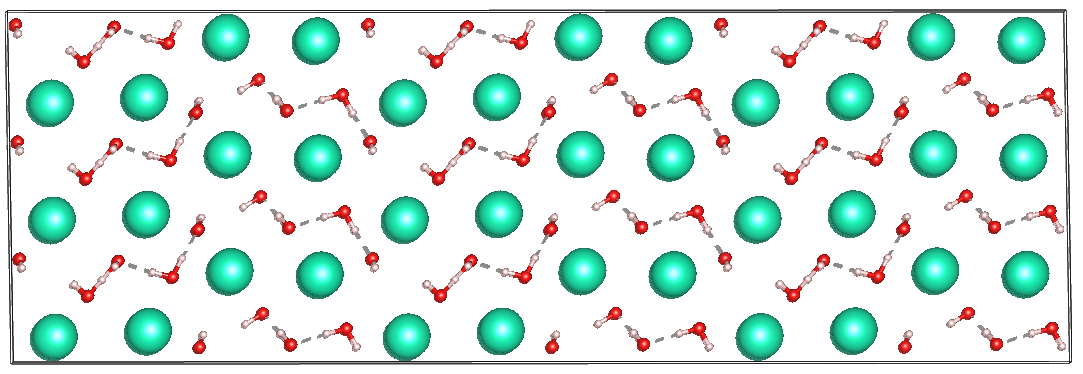}%
\caption{\label{I4amd}
Unstable I4$_1$amd structure with symmetry enforced  a=b=4.329, c= 14.25, and minimum energy P1 equivalent with symmetry relaxed and phonons frozen in a=4.619, b= 4.566, c=15.717.
}
\end{figure}

We set up a unit cell with the $I4_1/amd$ symmetry\cite{vcerny2002tetragonal} (fig.\ref{I4amd}a) and relaxed it (Table.\ref{tab:struc}).
In this structure each O has three neighbouring O with a hydrogen between. The formal symmetry places each hydrogen midway between two 
oxygen atoms, but in calculations this is highly unstable with respect to molecule formation.  Therefore we do not regard this as a credible representation of the bonding.

We then applied small random displacements to each atom removing the symmetry constraints, and re-relaxed.  This found the structure shown in  figure \ref{I4amd}b.  The Cs sublattice remains close to the original $I4_1/amd$ structure, but the hydrogens are off-centred, forming H$_2$O and OH units.  The symmetry is P1 (and probably non-unique), so there are a range of bondlengths.  The covalent bond in OH is significantly shorter than the H$_2$O one, while the hydrogen bond donated by OH is significantly longer.
 This symmetry-breaking reduces the energy of the ground state by about 0.6 eV/f.u. and increases the calculated volume of the zero-pressure structure by a surprisingly large amount (Table.\ref{tab:struc}). This suggests that despite the high energy, centred hydrogens might become more stable at very high pressure.

Although the high symmetry structure is unstable, it does serve as an estimate for the barrier for exchange of a hydrogen from H$_2$O to OH - if all 3 hydrogens are regarded as atop the barrier, it gives a barrier height of only 0.2 eV for each hydrogen.

\subsubsection{Hexagonal structure - 2D network}

An alternative set of structures are based on close-packed layers of Cs and oxygen occupying the interstitial sites.
It is helpful to consider first the high symmetry structure $P6/mmm$.  If one ignores the hydrogens this is  strukturbericht C32 (AlB$_2$-type).  The unit cell has just one formula unit: the Cs is arranged in a triangular lattice, the O atoms on a honeycomb lattice.  To preserve symmetry, the H is midway between O atoms.   All other structures can be regarded as symmetry-breaking from this.  

DFT calculation of this hypothetical ordered $P6/mmm$ shows it to be highly unstable with respect to molecule formation.  
The reported medium-temperature $P\overline{3}m1$ or $P{3}m1$ structure allows for molecule formation and a buckled H$_3$O$_2$-layer  structure, with reported partial-occupancy of the hydrogen sites in accordance with the ice rules: we were unable to find a stable, ordered molecular structure which preserves this symmetry.

The simplest stable molecular structure is a primitive unit cell with $Pm$ symmetry, around 0.5 eV/f.u. more stable than the symmetric structure.   While this is a substantial energy, it is an order of magnitude lower than the typical gas-phase formation energy of three covalent bonds, showing how weakly bound the water molecule are.  As with $I4_1/amd$ structure, the molecule formation is accompanied by a significant increase in volume.

Finally, the molecules can be allowed to move out of the plane and rerelaxed. This forms a buckled layer.  The OH and H$_2$O molecules tilt  without distorting the molecules\footnote{CASTEP eliminates the divergent dipole-dipole energy using an Ewald sum.}.
Curiously, relaxing the constraint converts the symmetry from $Pm$ to $Cm$ , along with an energy reduction of 66 meV/f.u. This single formula unit structure has the lowest energy we found.  

\begin{figure}
\includegraphics[width=0.49\linewidth]{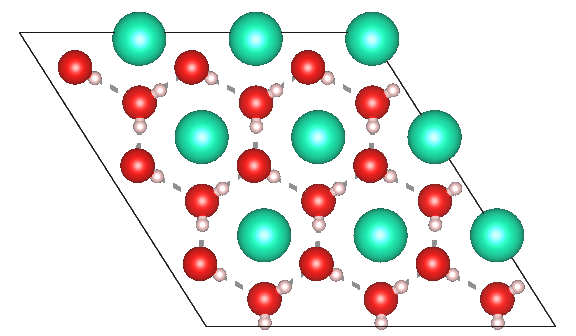}%
\includegraphics[width=0.49\linewidth]{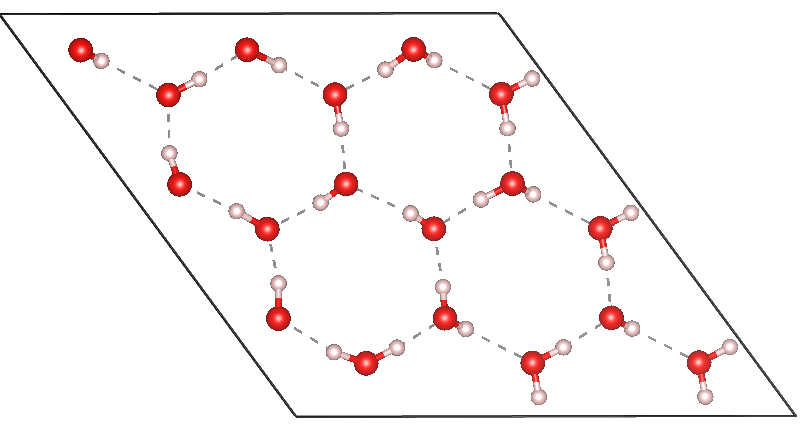}\\
\includegraphics[width=0.49\linewidth]{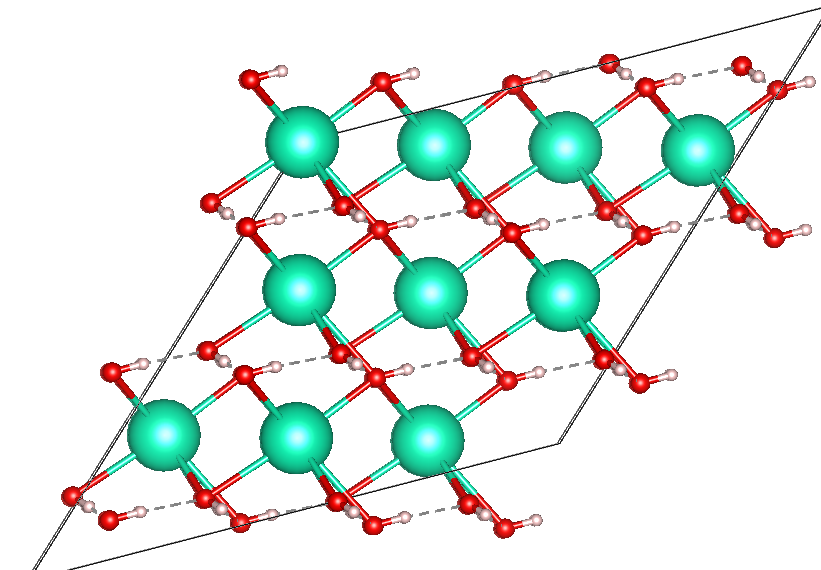}%
\includegraphics[width=0.49\linewidth]{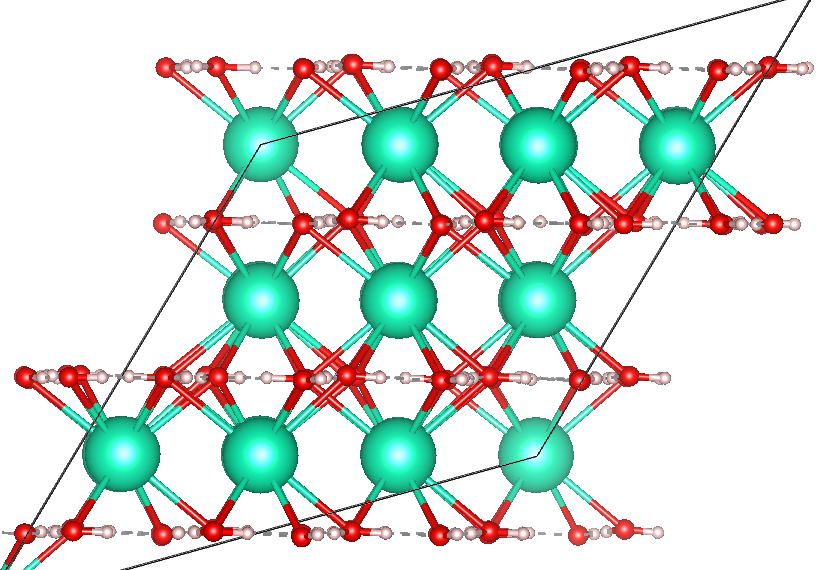}\\
\caption{\label{fig:layer} Top left: The arrangement of OH and H$_2$O molecules in the Cm structure.  The OH molecules  all point in the same direction.   Top right: Alternative bonding with disorder, showing an all-OH ring while still obeying the ice rules. 
Bottom left: Buckled layer from time averaged MD positions at 200K.
Bottom right: Flat layer from time averaged MD positions at 300K.}
\end{figure}

In the $Cm$ structure,  each  OH is hydrogen-bonded exclusively to H$_2$O and vice versa.  This high level of ordering follows from having a single unit cell.  The structure is monoclinic with all  molecular dipoles pointing in the same direction (Fig.\ref{fig:layer}).  

Even if the hydrogen bonding in each layer is fully specified, e.g. as in Fig.\ref{fig:layer}, there are six different variants of the layer, related by 60$^\circ$ rotations about a [001]-axis through the Cs. There are infinitely many ways of stacking such layers, and no guarantee that short-repeat ones will be stable\cite{loach2017stacking}.
We also tested structures in which different layers have OH molecules pointing in different directions. The simplest two-layer example has molecules in alternate layers rotated through 120 degrees $Cc$ symmetry and has the same energy as $Cm$ to within 1 meV/f.u. 
Another example has the OH in alternate layers pointing in opposite direction - which also requires the OH to be located directly above the H$_2$O.  This is an antiferroelectric ordering.  
By analogy with Ice VII-ice VIII  \cite{whalley1968ice,
jorgensen1983comparison,kuhs1986structure,kuhs1987geometry,haus1977model,ackland2025distinction}, one might expect that the antiferroelectric arrangement would be most stable, however the rotation of the OH, H$_2$O molecules out of the plane is already sufficient to eliminate the macroscopic dipole field, but again we find that the energy is indistinguishable from $Cm$ and $Cc$ stacking.

Finally, we investigated period-doubled structures which have hydrogen-bonds between pairs of waters and pairs of OHs. This does not break the ice rules but, assuming OH$^-$, it does bring like charges close together which will increase the energy.  This selection of structures is not exhaustive: the ice rules also allow for fully disordered structures which will have higher entropy and therefore may become stable at higher temperature.

 
 The H$_2$O acceptor/ OH donor hydrogen bond is significantly longer than the H$_2$O donor/ OH acceptor due to the unfavourable charge-dipole interaction.  There are no OH donor/ OH acceptor combination in the ordered structure: these would have even more unfavourable charge-charge interactions.

At increasing temperatures, disorder in both hydrogen position and H$_2$O-OH arrangement is likely to occur, possibly increasing the symmetry.   
These calculations establish the energetic hierarchy of disorder -
the ice rules and molecularisation are strong constraints, which imply equal numbers of OH and H$_2$O in each layer.  
 Ionic defects (H$_3$O or O$^{2-}$) or Bjerrum defects (O-H...H-O or O...O) are possible, but rare, costing hundreds of meV, including an out-of-plane molecular rotation to form intralayer defects. Alternating  OH - H$_2$O is the stable arrangement, but the cost of forming   OH-OH or  H$_2$O-H$_2$O  defects is in the 10's of meV.   The defect-defect interactions and interlayer correlation are very weak. 


\subsection{Molecular Dynamics}

We used Born-Oppenheimer Molecular dynamics on 180-atom supercells to study the dynamics of the hydrogen atoms. 
The 30 formula unit vicinal supercell has $a$ = 16.455\AA,  $b$ = 13.722\AA,  $c$ = 15.497\AA, $\alpha=\gamma=90$\textdegree, and $\beta  =   43.6$\textdegree. 
The main finding is that the water molecules break and reform on a picosecond timescale, with a hydrogen exchange reaction. 

\begin{equation} \rm {H_2O+OH \Leftrightarrow  OH+H_2O} \label{eq:reaction}\end{equation}

This reaction occurs continuously, leading to significant  disordering of the OH and H$_2$O from the idealised alternating arrangement.   As explained above, this local disorder does not imply a long-ranged disorder until the percolation threshhold is passed.   In table \ref{tab:struc} we outline the number of ``wrong" bonds - neighbouring pairs which are not one OH and one H$_2$O. In almost all cases, this means an OH-OH or H$_2$O-H$_2$O pair.  

These pairs are monitored dynamically based on the atomic positions.  At high temperatures it is possible that thermal fluctuation may result in false-positives for bond breaking where a hydrogen gets closer to its "unbonded" neighbour at the extremity of its oscillation, but return in the same vibration.  We tested this by quenching snapshots from the 600K simulations using the BFGS method\cite{fletcher2013practical}. This process sometimes eliminated ``wrong bonds". 
BFGS is not guaranteed to converge to the nearest minimum, it can undergo significant cooperative rearrangement; it is moot whether the dynamic or quenched count is the correct number of defects.

\begin{figure}
\includegraphics[width=0.49\linewidth]{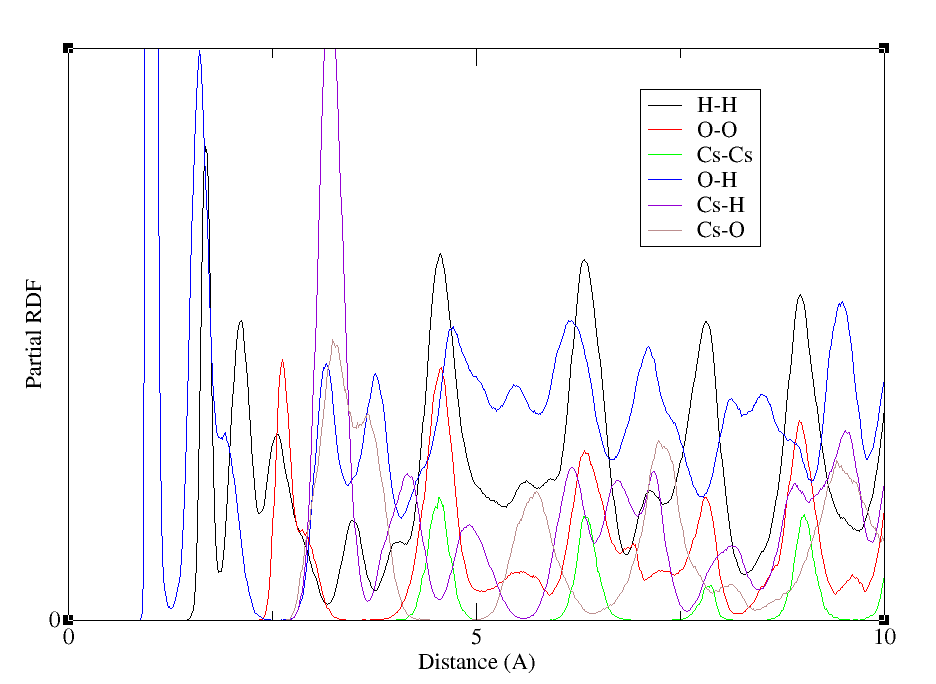} 
\includegraphics[width=0.49\linewidth]{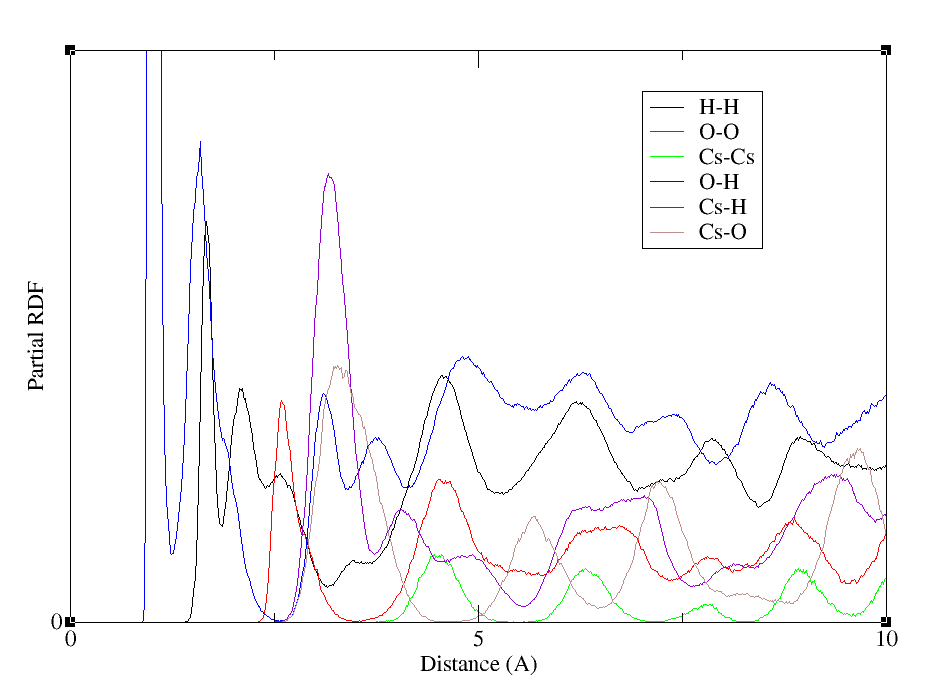} \\
\includegraphics[width=0.49\linewidth]{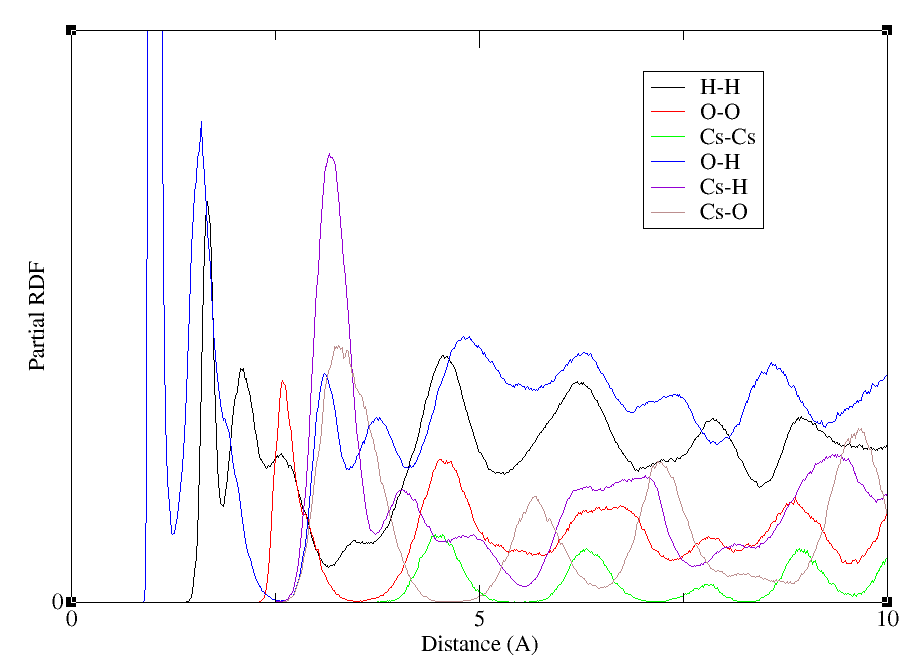} 
\includegraphics[width=0.49\linewidth]{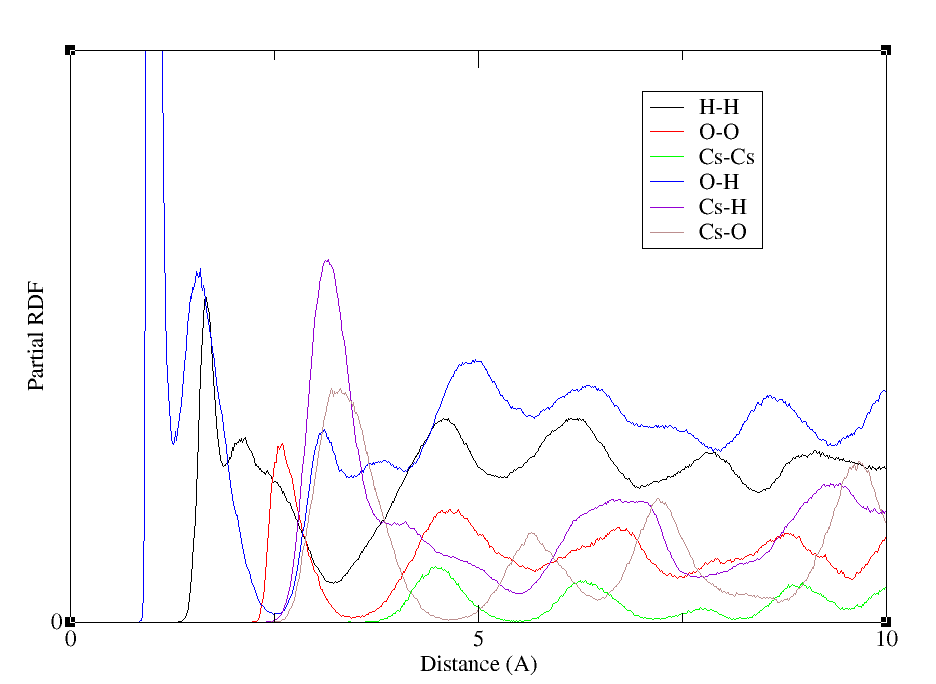} \\
\includegraphics[width=0.49\linewidth]{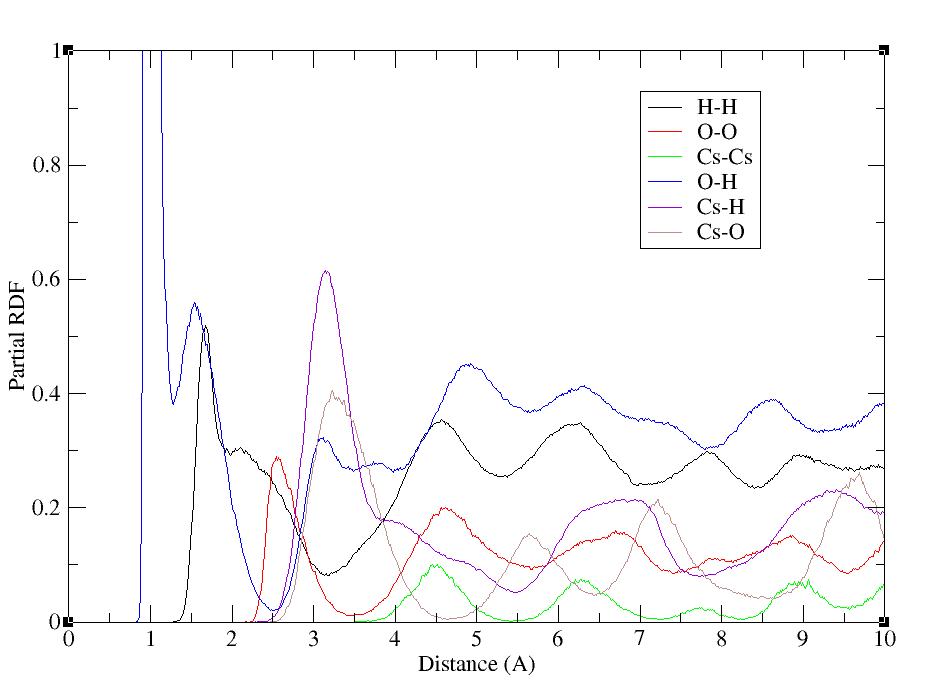} 
\includegraphics[width=0.49\linewidth]{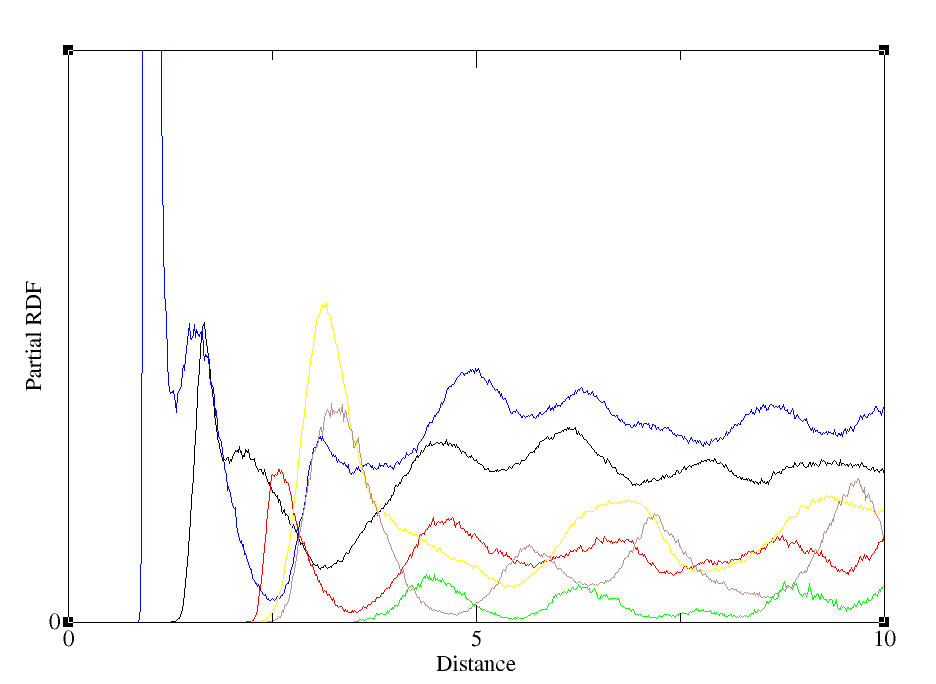} 
\caption{\label{fig:RDF}
Partial RDFs at temperatures 200K, 300K, 400K, 500K, 600K, and 700K.
}
\end{figure}

In figure \ref{fig:RDF} we show the partial radial distribution functions (RDF). These are all quite similar with temperature, showing that the change in long range order does not affect the short range order. All RDF peaks broaden and reduce in height with temperature, preserving coordination. 
The OH RDF shows the distinctive peaks associated with covalent and hydrogen bonds.  The peak at 1\AA\, contains the OH and H$_2$O covalent bonds, both longer on average than the typical values from vapour.  Detailed analysis, shows that the average H$_2$O bond is slightly shorter than the OH, but there is considerable overlap.  At all temperatures there is a distinct hydrogen-bond peak around 1.5\AA.
Other peaks are characteristic of the honeycomb structure with considerable hydrogen disorder.  The most striking trend is the second peak in the OH RDF which broadens significantly with temperature due to OH/H$_2$O disorder.

The mean squared displacements for each species  flatten off showing that the motion is oscillatory rather than diffusive.  Visualisation demonstrates that the Caesium ions remain on their lattice sites. The oxygen atoms show more displacement - in the $P6/mmm$ symmetry the OH plane is a mirror plane: in practice it is buckled with the oxygen partially occupying a site either above or below the mirror plane.  This disorder leads to the oxygen mean squared displacement (MSD) being significantly higher than Caesium.  The hydrogen RDF also plateaus, but at a significantly higher level: in addition to the buckling disorder,  each hydrogen is directly exchanging positions between two neighbouring oxygens
.
\subsection{Dynamical network analysis}
To follow the evolution of the structure we associated each hydrogen with precisely two oxygens, and assumed a covalent bond to the nearest oxygen and a hydrogen bond to the second nearest.  This forms an undirected network with oxygen ``nodes" and hydrogen ``links", which can in turn be represented by an adjacency matrix.  This adjacency matrix turns out to be constant throughout the simulation at all temperatures: consistent with the MSD, there is no exchange of oxygen atoms between sites.

We also construct a directed adjacency matrix, again with nodes representing oxygen and links representing hydrogen. 
The direction is towards the covalently bonded site.  This allows us to define, in principle, O, OH, OH$_2$ and OH$_3$ molecules.  Using those labels, we can determine the number of OH-OH and H$_2$O-H$_2$O nearest neighbours (Table \ref{tab:MD}).  This appears to go through two distinct transitions.  At 200K there are very few misbonds and the time-averaged OH layer remains buckled.  At 300K, the number of misbonds has increased by an order of magnitude.  Individual snapshots show buckled OH layers, but the time-average is flat. Another jump in misbonds occurs between 400K and 500K: this is more driven by the longevity of the misbonds than their frequency of production.

The number of OH and H$_2$O pairs fluctuated significantly during the simulation.

\subsection{Percolation Transition: OH-H$_2$O ordering}

Percolation theory shows that for a 2D arrangement honeycomb arrangement the bond percolation threshold\cite{sykes1964exact},  at which an infinite cluster of connected bonds forms, is $p_{c}\approx 0.652$.  If we define a ``bond" for OH-H$_2$O pairs, then the percolation threshold is where long range site ordering of the alternating OH and H$_2$O breaks down, and the associated diffraction signal will vanish.  
We believe that percolation is the most reliable way to detect a transition due to loss of long-ranged order in our simulations, because direct measurement of long range order is hindered by the periodic boundary conditions.

In our system we sample 60 sites, leading to 90 bonds, so under the assumption that the wrong-bonds are uncorrelated, the disorder transition occurs when there are 90(1-p$_c$)=31.3  OH-OH or H$_2$O-H$_2$O pairs.  Table.\ref{tab:MD} shows that this number of wrong bonds is not reached. Since the loss of long ranged order occurs before the percolation transition, it implies that the wrong-bonds are not randomly placed. This will be driven by the ice-rules and OH$^-$ charges, which reduce clustering of OH, and the finite size of the simulation cell which tends to promote ordering.

\subsection{Hydrogen diffusion} 
We see no diffusive behaviour, nor in-plane rotation of individual molecules. This is unsurprising because 120$^\circ$  in-plane rotation of an individual molecule would violate the ice rules.   

However, we do observe a thermally-activated mechanism which would enable it.  A water molecule can rotate out of the plane to form a bond to the layer above, creating a link in the third dimension.  This creates a vacancy in the 2D honeycomb layer which can then move  in a similar way that KOH is used to facilitate transitions in ice by introducing vacancies  (L-defects)\cite{Tajima1982,salzmann2006preparation}.  In practice, dopants may also contribute to the observed ionic conductivity, but our simulations reveal that there is an intrinsic mechanism even in stoichiometric CsOH$\cdot$H$_2$O.

\subsection{Microstate energies and Raman}
The MD allows us to consider a number of microstates with different ordered H-positions, and with different orderings of OH and H$_2$O molecules.  We  generated other structures by quenching snapshots from the 600K MD simulations with BFGS and correlating those against the number of OH-OH near neighbours, a proxy for the species disorder. Interestingly, some 30\% of the OH-OH defects identified in the high temperature MD are eliminated in the relaxation process.  We find a clear correlation between the number of defects and the enthalpy.    

Routine Raman calculations are complicated by the need to use ordered supercells, so that unique calculation for the disorder material is impossible.  We carried out calculations using DFPT on a variety of model structures and identified strong Raman bands corresponding to the intramolecular stretching modes of OH in the range 3350-3550cm$^{-1}$ and of H$_2$O in the range 2550-2750cm$^{-1}$.  H$_2$O bending modes lie between 1400-1700cm$^{-1}$, and a wide range of librations between 650 and 1100cm$^{-1}$.  Translational modes in the OH plane are around 200-300cm$^{-1}$ with layer modes at 100cm$^{-1}$ and below.  These calculations are consistent with experimental work\cite{Lutz1988} and show that broad Raman bands arise from diverse local environments as well as thermal and lifetime broadening.

\section{Raman signal from a tunnelling hydrogen}

The proton exchange  reaction (Eq. \ref{eq:reaction}) is possible only for an H$_2$O donating to an OH.
 In other cases, e.g. OH donor or H$_2$O donor to another H$_2$O, exchange is impossible and a standard OH stretch is observed.

We predict that the combination of vibration and exchange will give a complex Raman signature, which we model using a one–dimensional Hamiltonian of the form
\begin{equation}
    \hat{H} = -\frac{\hbar^{2}}{2m}\frac{d^{2}}{dx^{2}} + V(x),
\end{equation}
where $m$ is the effective mass associated with the vibrational coordinate.  
The potential energy surface consists of a harmonic well perturbed by a localized Gaussian term,
\begin{equation}
    V(x) = \frac{1}{2} m \omega^{2}(x-x_{0})^{2}
           + A \exp\!\left[-\frac{(x-x_{b})^{2}}{2\sigma^{2}}\right],
\end{equation}

Parameters ($\omega$, $A$, $x_{b}$, $\sigma$) are chosen to yield an energy surface similar to the observed motion of one hydrogen atom between oxygens.  Notably, there are two inequivalent minima corresponding to the position which fits the water-hydroxyl ordering and the one which breaks it. The barrier and the relative depths of the minima depend on the environment of the OH-H$_2$O pair,  but can be estimated from the DFT calculations\cite{gao2016symmetric}.

The potential $V(x)$ was discretized on a uniform real–space grid $x_{i}$, and the 1-D Schroedinger equation was solved numerically.    

To facilitate comparison with experiment,  we calculate the  vibrational level spacings in cm$^{-1}$.
Raman intensities were computed using a linear model for the polarizability,
\begin{equation}
    \alpha(x) = \alpha' x 
\end{equation}
which captures the linear contribution to the coordinate dependence of the electronic response.   
The Raman transition strength for $i\rightarrow j$ is proportional to the square of the polarizability matrix element,
\begin{equation}
    S_{i\rightarrow j} = 
    \left| \langle \psi_{j} | \alpha(x) | \psi_{i} \rangle \right|^{2},
\end{equation}
and the temperature dependence is introduced through Boltzmann populations,
\begin{equation}
    P_{i}(T) = 
    \frac{\exp[-E_{i}/(k_{\mathrm{B}}T)]}
         {\sum_{k}\exp[-E_{k}/(k_{\mathrm{B}}T)]}.
\end{equation}
The Raman active modes are
\begin{equation}
    I(\tilde{\nu};T) = 
    \sum_{i,j} P_{i}(T)\, S_{i\rightarrow j}\,
    \delta\!\left(\tilde{\nu} - \frac{E_{j}-E_{i}}{hc}\right).
\end{equation}
To model finite experimental resolution and homogeneous broadening, each transition was convoluted with a Lorentzian of half–width $\gamma$,
\begin{equation}
    L(\tilde{\nu};\tilde{\nu}_{0},\gamma)
    = \frac{1}{\pi}\frac{\gamma}{(\tilde{\nu}-\tilde{\nu}_{0})^{2}+\gamma^{2}}.
\end{equation}

This anharmonic potential gives different Raman shifts for different $m,n$ combinations with very different character.   For excitation from the ground state, $0\rightarrow 1$ we find a particularly high frequency peak corresponding to the stretching mode of the H$_2$O covalent bond. Then, for $1\rightarrow 2$, we find a low frequency mode corresponding to a reaction  OH+H$_2$O$\rightarrow$ H$_2$O+ OH.  Although this is locally a symmetry hydrogen exchange reaction, the neighbouring ordering of OH and H$_2$O is affected so there is an energy change as well as a barrier, and this also varies from site to site.   The intensity of the Raman peak associated with the OH+H$_2$O$\rightarrow$ H$_2$O+ OH transition is proportional to the occupation of the $\nu=1$ level, so it manifests as a low energy peak appearing at elevated temperatures.

\begin{figure}
\includegraphics[width=0.49\linewidth]{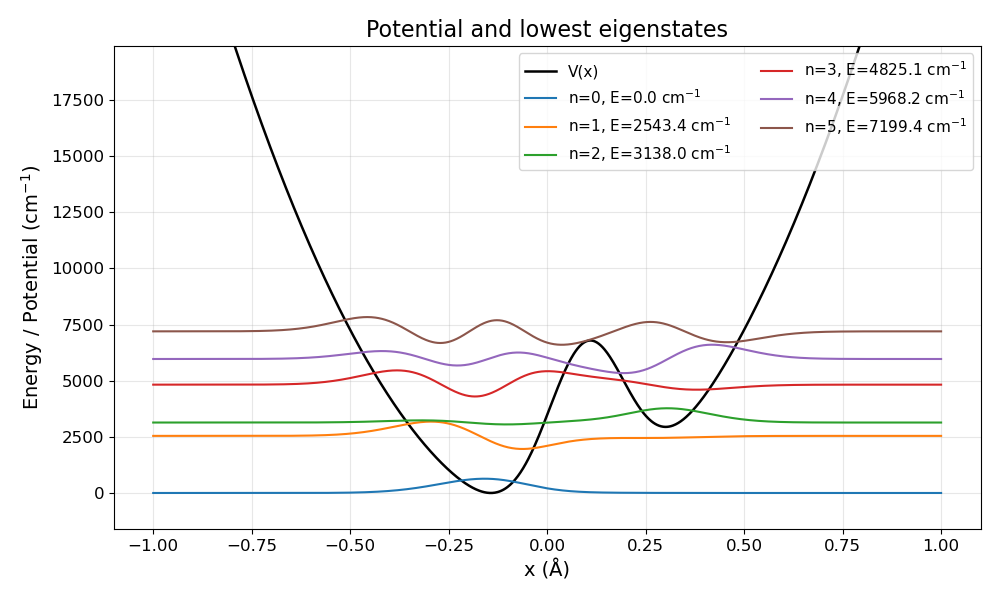} 
\includegraphics[width=0.49\linewidth]{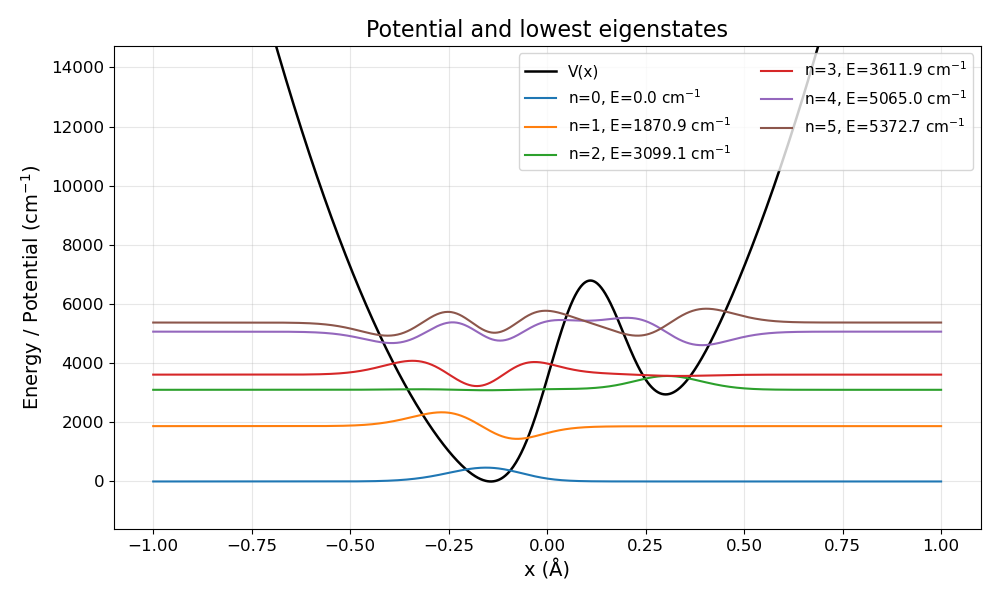} \\
\includegraphics[width=0.49\linewidth]{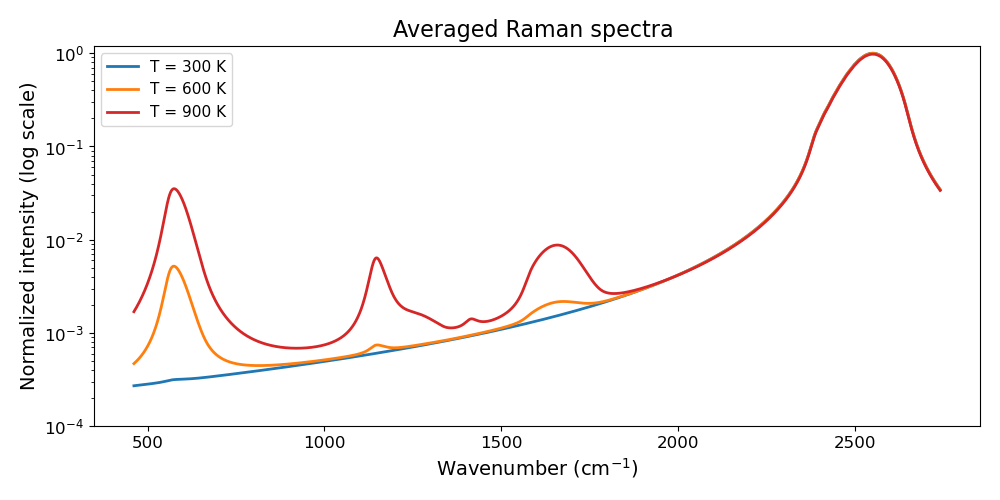} 
\includegraphics[width=0.49\linewidth]{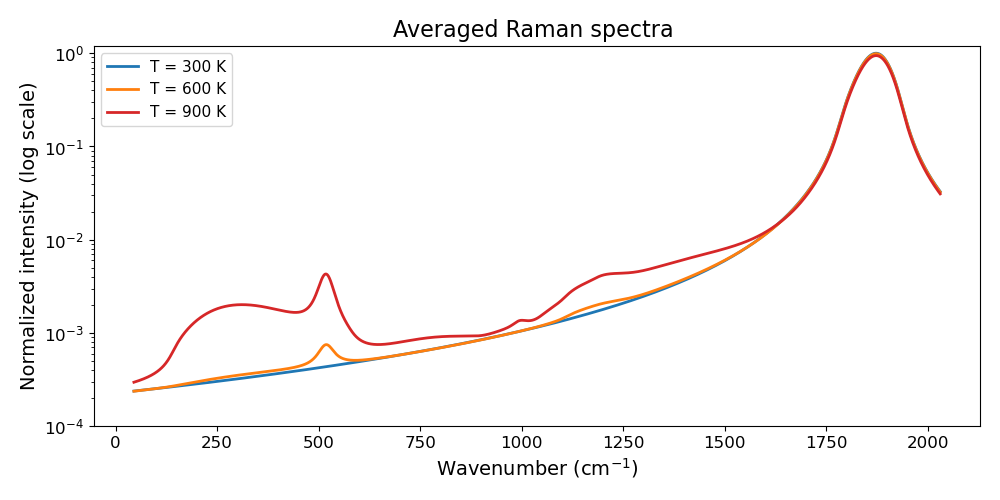}
\caption{\label{fig:anhamonic} Anharmonic potential model for H$_2$O..OH mode. Top: potential and the associated wavefunctions for hydrogen and deuterium.  For these parameters, the ground state (blue) clearly corresponds to the hydrogen in the leftmost potential well: an OH bond.  The first excited state can be regarded not as an excitation of this OH stretch, but rather the hydrogen bonded to the other hydrogen.  The second excited state is closer to an excitation of the OH stretch, Bottom: Contribution to Raman spectrum from this mode only, at a range of temperatures, using a range of bump heights and fundamental frequencies from 2391-2641 to represent the range of environments of the water donors and OH acceptors. 
The transitions $\nu= 1\rightarrow 2$ and  $\nu= 1\rightarrow 3$ become evident at higher temperature, when the  $\nu= 1$ mode is thermally occupied.
The parameterisation here shows the qualitative behaviour: the temperature and isotope effects.  It is not intended to be compared quantitatively to experimental values}
\end{figure}

The different characters of the transitions leads to dramatic isotopic differences in the Raman signal of this mode.   In Fig.\ref{fig:anhamonic} we show energy levels corresponding to H or D for illustrative potential $V(x)$.  In each case, the first excited state behaves like a quantum oscillator giving peaks differing by about $\sqrt{2}$.  The second excited state is centred on the other potential well, and has approximately the same energy for both H and D.  We averaged over a credible range of barrier  parameters for $V(x)$ and calculated the Raman spectrum.  At 300K, only the fundamental mode is visible,  but at higher temperature other transitions become significant in the hydrogen spectrum.


\begin{figure}
\includegraphics[width=\linewidth]{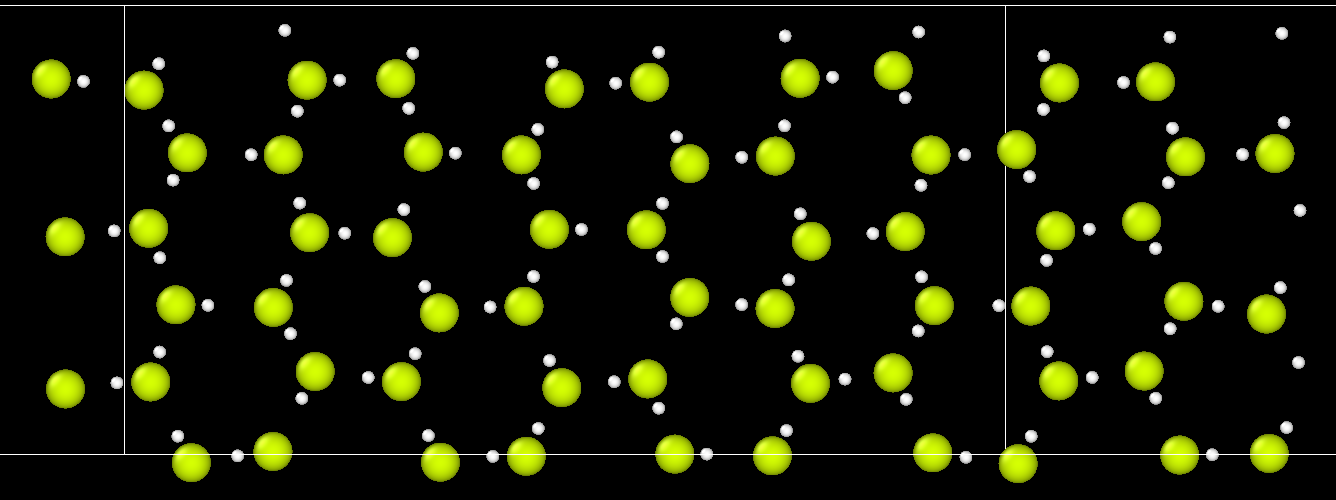}\\
\includegraphics[width=\linewidth]{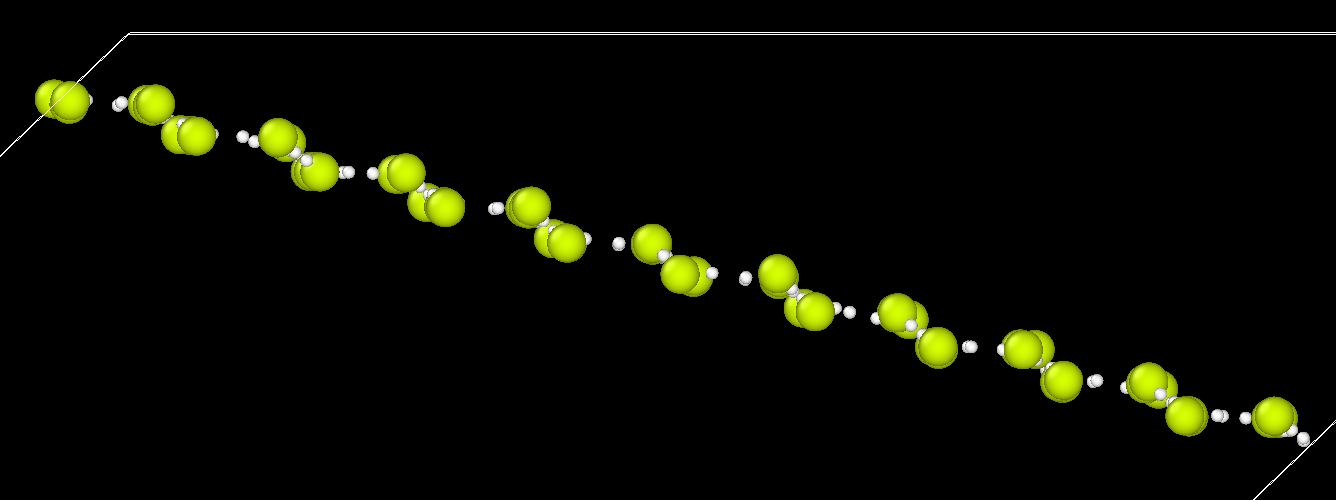}%
\caption{\label{fig:md}Illustrating the vicinal molecular dynamics cell.
Top figure shows all 60/120 independent O/H atoms as a single layer.  Lower figure shows side view of the layer in $Cm$, emphasising the buckling.  This structure was quenched from a 600K MD snapshot.  It can be seen that the ice rules are followed, all oxygens can be assigned to OH or H$_2$O molecules but the OH and H$_2$O are somewhat disordered.  Complete disorder is prohibited by the ice rules: it is possible to have an OH-only 6-membered ring, but two adjacent rings must contain at least one OH and one H$_2$O. Cs atoms are omitted for clarity.
} 
\end{figure}

\begin{table*}
\begin{tabular}{cccccc}
T&  timesteps & transitions & misbond (t) & misbond \% &Mean OH
\\ \hline
200K&  15078 & 174 & 10470 &  0.695 & 1.749 (0.207)\\
300K& 15251 & 1248  &128094 & 8.40 &  1.739 (0.240) \\
 400K& 23754 & 3979  & 168197& 9.28 &1.738 (0.254)\\
 500K& 14247  &4604 & 251027 & 17.62 & 1.743 (0.279) \\
 600K& 11314 & 5547 & 261966 & 23.15 & 1.760 (0.303) \\
 700K& 9000 & 4314 & 217500 & 24.17 & 1.803 (0.343) \\
\end{tabular}\label{tab:MD}
\end{table*}

\begin{table*}
\label{tab:struc}
\caption{Structural information from the rSCAN calculations: lattice parameters, covalent and hydrogen bondlength labelled by donors.  The symmetry enforced $I4_1/amd$ and $P6/mmm$ structures have centred hydrogens rather than molecules and are extremely unstable, although their higher density suggests they may become stable at extreme pressure.}
\begin{tabular}{c|cc|ccc}
& $I4_1/amd$ & $P1 (I4_1/amd)$ &  $P6/mmm$ &  $P3m1$ & $Cm$ \\ \hline
c (\AA) & 14.25&15.577 &4.455 &4.366  & 4.434\\
a (\AA) &4.329& 4.520,4.464 & 4.206& 4.631,4.575 & 4.563\\
V (\AA$^3$/fu) &66.78 & 78.55 & 68.25& 77.84 & 77.85 \\
r$_{OH}$(\AA)& 1.219& 0.966-0.968 &1.214&0.967&0.967 \\
r$_{H_2O}$(\AA)&1.220&  1.011-1.017 & 1.214 &1.011, 1.014 & 1.014\\
d$_{OH}$(\AA)&1.220& 1.93-1.99 & 1.214 & 2.001 & 1.951 \\
d$_{H_2O}$(\AA)&1.219& 1.57-1.62 & 1.214&1.588,1.598 & 1.577\\
\end{tabular}
\end{table*}


\section{Discussion and Conclusions}

In this paper we have investigated the behaviour of hexagonal Caesium hydroxide hydrate: CsOH$\cdot$H$_2$O, which has a hydrogen-bonded 2D layered structure, the two dimensional analogue of water.  We demonstrate that the OH and H$_2$O molecular units are well-defined at all temperatures up to 700K, however the individual molecules are extremely short-lived, with lifetimes on the picosecond scale.  The ice-rules are obeyed at all temperatures, the only anomaly being rare out-of-plane rotations at high temperatures which form short-lived bonds connecting layers.  

The experimental structures can be understood from MD by considering the ordering. 
At high T we find disordered OH and H$_2$O, giving an average structure with $P6/mmm$ symmetry,  On cooling we obtain long range ordering of the two species giving $P\bar{3}m1$, and buckling of the OH layers breaks inversion symmetry to $P{3}m1$.  Finally, the molecular orientations become fixed forming the monoclinic $Cm$ structure\cite{jacobs1982struktur}.

There is a general preference for OH and H$_2$O molecules to be ordered on adjacent sites, with the number of  OH-OH and H$_2$O-H$_2$O pairs increasing with temperature.

We find no evidence for superionic (superprotonic) behaviour, in contrast to the title of a recent paper\cite{rodenburg2025superprotonic}.  However, there is no conflict with the high proton conductivity reported there which is readily explained by rapid hopping of defects in off-stoichiometric compounds or through the self-generated vacancy mechanism associated with out-of plane rotation

The H$_2$O - OH donor case leads to an unusual set of excited states combining OH stretching and H-exchange.  Exact details are sensitive to the exact parameterisation, but the distinctive features are the appearance of new Raman peaks at high temperature, and isotopic differences in both lineshape and frequencies very different from the simple $\sqrt{2}$ scaling expected for harmonic oscillators.

\begin{acknowledgments}
We  acknowledge EPSRC for support under grant No. EP/Y020987.
\end{acknowledgments}

\bibliography{apssamp,forCsOH_H2O}

\end{document}